\input{aipcheck}

\documentclass[
    ,final            
  ]
  {aipproc}

\layoutstyle{8x11double}

\begin{document}

\title{Searches for the Most Metal-Poor Candidates from \\
SDSS and SEGUE}



\author{Timothy C. Beers}{
  address={Department of Physics \& Astronomy, CSCE: Center for Study of
Cosmic Evolution, and JINA: Joint Institute for Nuclear Astrophysics, Michigan
State University, E. Lansing, MI 48824-1116, USA}
}

\author{Daniela Carollo}{
  address={Osservatorio Astronomico di Torino, Italy, and JINA: Joint Institute for Nuclear Astrophysics, Michigan
State University, E. Lansing, MI 48824-1116, USA}
}


\begin{abstract}

We report on efforts to identify large samples of very and extremely metal-poor
stars based on medium-resolution spectroscopy and $ugriz$ photometry obtained
during the course of the Sloan Digital Sky Survey (SDSS), and its extension,
SDSS-II, which includes the program SEGUE: Sloan Extension for Galactic
Understanding and Exploration. To date, over 8000 stars with [Fe/H] $\le -2.0$
and effective temperatures in the range 4500~K $ < T_{\rm eff} < 7000$~K have
been found, with the expected numbers in this temperature range
to be well over 10,000 once SEGUE is completed. The numbers roughly double when
one includes warmer blue stragglers and Blue Horizontal-Branch (BHB) stars in these
counts. We show the observed low-metallicity tails of the Metallicity
Distribution Functions for the cooler SDSS/SEGUE stars obtained thus far. We
also comment on the confirmation of an inner/outer halo dichotomy in the Milky
Way, and on how this realization may be used to direct searches for even more
metal-poor stars in the near future. 

\end{abstract}

\maketitle

\section{Introduction}

The primary recent searches for metal-poor stars in the Galaxy (the HK survey of
Beers and colleagues [1], and the Hamburg/ESO Survey of Christlieb and
colleagues [2]) were based on objective-prism techniques, followed by many years
of follow-up medium-resolution spectroscopy with intermediate-aperture
telescopes. Although this approach has been spectacularly successful, and has
resulted in the identification of several thousand stars with [Fe/H] $< -2.0$
(VMP stars according to [3]), and several hundred stars with [Fe/H] $< -3.0$
(EMP stars according to [3]), only three stars have been found to date with
[Fe/H] $< -4.0$ (UMP according to [3]), several of which are discussed in detail
by other contributions in this volume. 

Clearly, much has already been learned from high-resolution spectroscopic
studies of many of the HK and HES stars, but many questions concerning the
nature of the first generations of objects born in the Galaxy
remain -- questions that can only be answered by obtaining a significant
increase in the numbers of VMP, EMP, and especially UMP and HMP ([Fe/H] $< -5.0$
according to [3]) stars. New methods and new strategies are required. Here we
report on one such effort, based on observations obtained during the SDSS and
SEGUE [4]. Other survey methods and approaches are discussed by Christlieb et
al. in this volume.  

Kinematic studies of the SDSS stellar calibration objects, by Carollo et al. [5],
has confirmed the (long suspected) dichotomy of the halo of the Milky Way. Aside
from the obvious importance of this result for understanding the assembly and
evolution history of the halo itself, the fact that the outer-halo population
appears to exhibit a Metallicity Distribution Function (MDF) that peaks around
[Fe/H] $= -2.2$, roughly 0.5 dex lower than that of the inner-halo population,
opens up the possibility of specifically targeting the lowest metallicity stars
in the Galaxy {\it if one could efficiently pre-select} likely members
of the outer halo, as is discussed in more detail below.

\section{Metal-Poor Stars Discovered by the SDSS and SEGUE Surveys}

Metal-poor stars are identified from several sources within the SDSS and SEGUE.
In the original SDSS, numerous metal-poor stars are found among the 16 (color-
and apparent-magnitude selected) spectrophotometric and telluric calibration
stars that are targeted each time an exposure of a spectroscopic plug-plate
(containing a total of 640 fibers) is obtained. In addition, QSO candidates that
turn out not to be quasars, and BHB candidates that turn out to be somewhat
cooler than expected, based on spectroscopic follow-up, are often
low-metallicity stars. Additional metal-poor stars are found among the
calibration objects being observed during the ongoing LEGACY spectroscopy
program. The continuing SEGUE program specifically targets metal-poor stars, among
the 13 or so categories it considers. There are also many metal-poor stars that
are found among the F turnoff, G dwarf, and K giant target categories. As in
SDSS, each SEGUE plug-plate obtains 16 calibration stars as well. The total
number of stars observed spectroscopically by either SDSS or SEGUE is currently
over 350,000. It should be noted that the color selection, either for the
calibration objects or the low-metallicity candidates, is not capable of
discriminating between stars of metallicity lower than [Fe/H] = $-2.0$, since
the effect of declining metallicity on broadband stellar colors is minimal in
this regime. Hence, the MDF of such stars should be relatively unbiased below
this metallicity.

Stellar atmospheric parameters ($T_{\rm eff}$, log g, and [Fe/H]) are determined
for SDSS/SEGUE stars by application of the SEGUE Stellar Parameter Pipeline
(SSPP), which is described in detail by Lee at al. [6,7]. Tests of the SSPP
indicate that, over the temperature range 4500 $\le T_{\rm eff} \le 7500$~K,
external accuracies in the derived parameters are on the order 120~K, 0.25 dex,
and 0.2 dex, for $T_{\rm eff}$, log g, and [Fe/H], respectively.

Figure 1 shows the MDF for the over 4200 stars from SDSS with adopted [Fe/H]
$\le -2.0$ (and with $S/N \ge 10/1$). Figure 2 shows the same distribution for the
over 2400 similar stars observed during the first year of SEGUE. The addition of
observations from LEGACY and from the second year of SEGUE increases the total
numbers of stars with [Fe/H] $\le -2.0$ to well over 8000 objects in the
temperature range considered herein. This already
represents roughly a tripling of the total numbers of VMP stars obtained
by the sum of all previous surveys conducted in the past half century since the
low-metallicity star phenomenon was recognized. 

\begin{figure}[!t]
\includegraphics[height=2.75in]{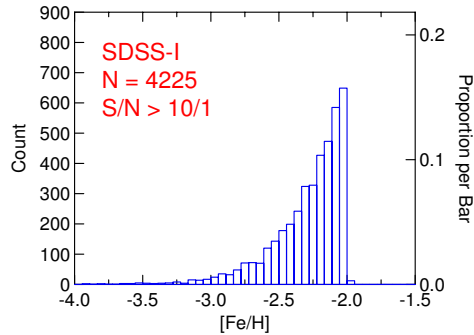}
\caption{The MDF for VMP stars with 4500~K $ < T_{\rm eff} < 7000$~K 
observed during the course of SDSS-I.  The great majority of these stars 
are objects taken for spectrophotometric and telluric calibrations.}
\end{figure}

\begin{figure}[!t]
\includegraphics[height=2.75in]{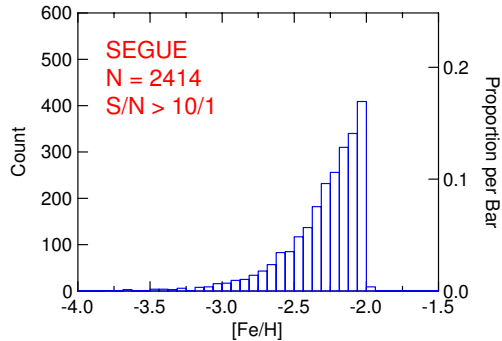}
\caption{The MDF for VMP stars with 4500~K $ < T_{\rm eff} < 7000$~K observed 
during the course of the first year of SEGUE.  These
stars include objects specifically targeted to be likely VMP stars, as well as
calibration stars.}
\end{figure}

It is clear from the shapes of the MDFs in Figures 1 and 2 that
the total number of EMP stars (with [Fe/H] $\le -3.0$) is declining very
rapidly, with only a handful of stars being identified to date with [Fe/H] $\sim
-4.0$, and none below this value.  We address the reasons for this below.  

\begin{figure}[!t]
\includegraphics[scale=0.7]{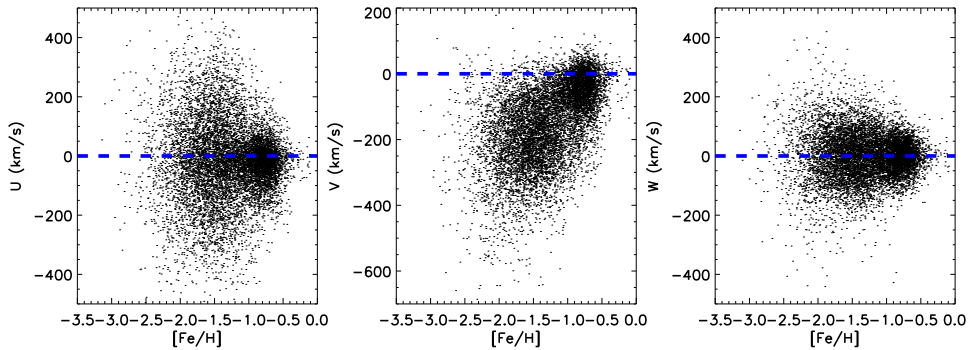}
\caption{The components of space motion for the local sample of stars in 
Carollo et al., shown as a function of [Fe/H].  The dashed lines indicate
the location of the Local Standard of Rest (LSR).  As can be seen in the middle panel,
the thick-disk and metal-weak thick disk population exhibits a small lag with
respect to the LSR.  Note also, in this panel, the large numbers of VMP stars
with large retrograde motions ($ V < -220$ km/s) with respect to the center of
the Galaxy.}
\end{figure}

\section{The Dichotomy of the Galactic Halo}

The region of the Milky Way halo, beyond the thin- and thick-disk systems, has long
been thought of as a single entity, comprising old stars and globular clusters
that represent the earliest populations of objects to have formed in our Galaxy.
Although many previous studies have speculated that the halo may in fact be
partitioned into inner and outer components, possibly with different kinematics
and spatial distributions, these studies have been limited by the generally
small numbers of stars or globular clusters upon which this split could be made.
Hence, it has always remained a suspicion, not a certainty.  Until now.
Based on medium-resolution spectroscopy for an initial sample of 20,366 SDSS
calibration stars located within roughly 20 kpc of the Sun, Carollo et al. [5] have
shown that the halo is clearly divisible into two broadly overlapping structural
components. These are: (1) the inner halo, which is dominated by stars on highly
eccentric orbits and exhibits a peak metallicity [Fe/H] $= -1.6$, as well as a
flattened density distribution with a modest net prograde rotation, and (2) the
outer halo, which includes stars that possess a wide range of orbital
eccentricities (including many on low eccentricity orbits), exhibits a peak
metallicity [Fe/H] $= -2.2$, and a spherical density distribution with a highly
statistically significant net retrograde rotation (between $V_{\phi}$ = $-50$ and
$-80$ km/s, depending on which subset of the data is considered).

The kinematic analysis of the Carollo et al. data set is carried out for the
subset of over 11,000 stars located within 4 kpc of the Sun, where one can take
advantage of the existence of reasonably accurate proper motions obtained from a
re-calibration of the USNO-B catalog, as described by Munn et al. [8]. Typical
errors in the derived proper motions are on the order of 2.5 to 3.5 mas/year,
which is more than adequate for this exercise.  Figure 3 shows the distribution
of derived $UVW$ components of the space motions for these stars as a function of
[Fe/H], as determined by the SSPP.  The impact of the size of this sample can be
appreciated by comparing with similar diagrams in the article by Frebel et al.
in this volume.  

As is clear from inspection of Figure 3, one can clearly make out the presence of
the concentrations of stars with relatively low space motions and moderately high
metallicities, which we associate with membership in the thick-disk and metal-weak
thick-disk populations.  The very large velocity dispersions at lower
metallicity are of course associated with the halo population(s) (which do not
clearly separate from one another in such diagrams).   

Stars in the disk population(s) can be effectively eliminated by considering
only those stars on retrograde orbits. Furthermore, as one takes subsets of the
stars that achieve maximum distances above the plane in the course of their
orbits about the center of the Galaxy (Z$_{\rm max}$) of 5 kpc or more, one can
address whether or not they exhibit properties consistent with that expected
from a single, or a more complex, halo population. Figure 4 shows the distribution
of [Fe/H] for stars with varying levels of increasingly retrograde orbits, and
with different cuts on Z$_{\rm max}$. As is clear from inspection of this
Figure, as one sweeps to more retrograde orbits, and to greater Z$_{\rm max}$,
the nature of the MDF changes over to favor stars at lower [Fe/H], and exhibits
fewer stars with higher [Fe/H]. This is due to the change in the relative
dominance of the outer-halo population over the inner-halo population at low
metallicity and at distances greater than roughly 15-20 kpc from the Galactic
center. This dichotomy may have a profound influence on searches for the most
metal-poor stars in the Galaxy in the future, as described below.

\begin{figure}
\includegraphics[scale=1.0] {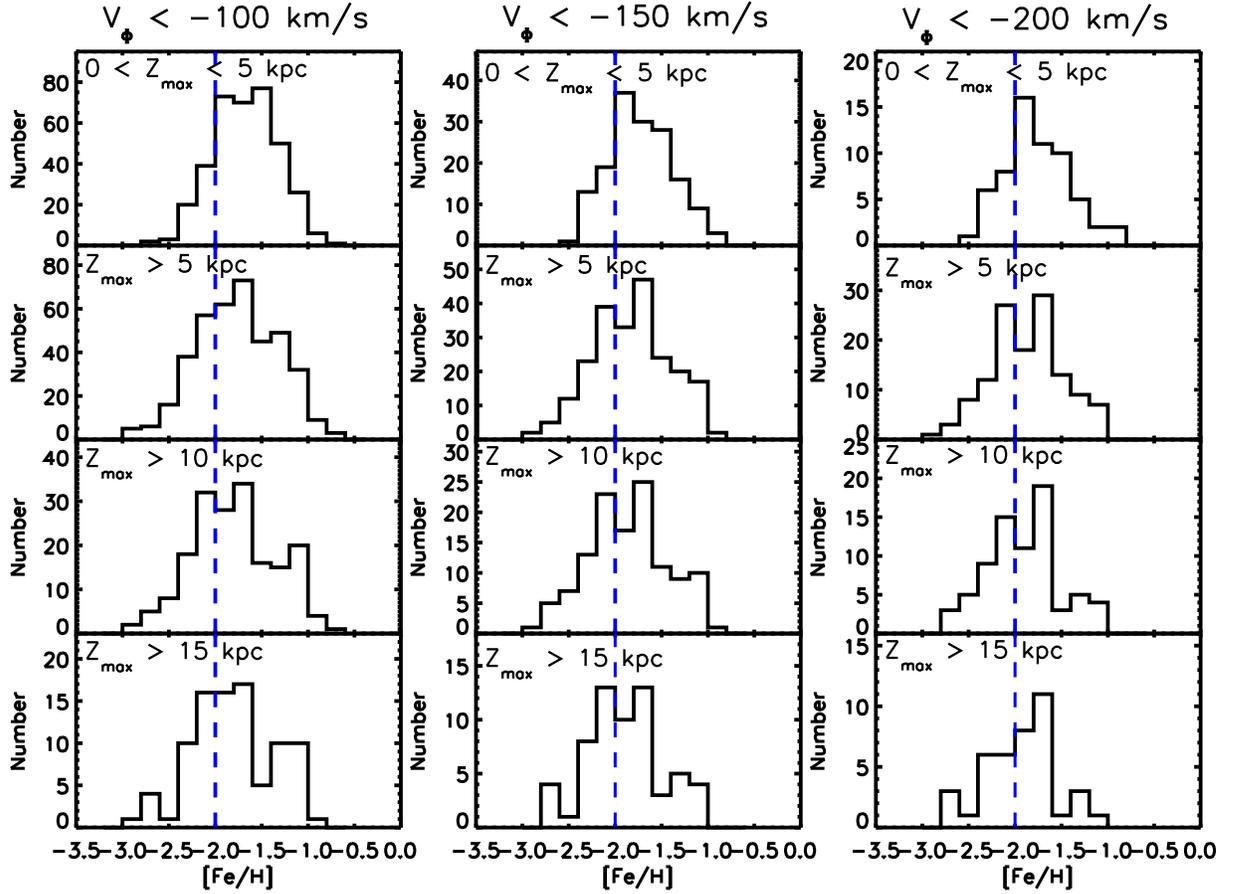}
\caption{Plots of the MDFs for stars on highly retrograde orbits with respect to the
Galactic center.  The left hand column corresponds to stars with $V_{\phi} < -100$
km/s, while the middle and right-hand columns correspond to stars with more extreme
retrograde motions, $V_{\phi} < -150$ km/s and $V_{\phi} < -200$ km/s,
respectively.  From top to bottom in each column, the panels show the MDFs for
stars with increasing cuts on Z$_{\rm max}$.  The vertical lines at [Fe/H]$ =
-2.0$ are shown for reference. Note that the nature of the MDFs changes within
each column, gradually favoring stars with [Fe/H] $< -2.0$ at higher Z$_{\rm
max}$. This change is even more evident (even though the total numbers of stars
are fewer) for the most extremely retrograde columns. All of the cuts on
individual panels with Z$_{\rm max} > 5$ kpc are highly inconsistent, according to a K-S
test, with being drawn from the same parent MDF as that of the low Z$_{\rm
max}$ cuts shown in the upper panels.}
\end{figure}

\section{Future Searches for the Most Metal-Poor Stars}

There are two primary techniques that have been exploited to date for
the identification of low-metallicity stars in the Galaxy.  The first
used selection of likely halo members based on observations of stars in the
local volume of the Galaxy with relatively high proper motions, e.g., Ryan and
Norris [9], or Carney et al. [10]. The second was based on the identification of
stars from $in-situ$ surveys (such as the HK survey and the HES) that sought to
find halo members based on low-resolution objective-prism spectroscopy to a
limiting magnitude between $B \sim 15.5$ (HK survey) and $B \sim 17.5$ (HES).
The SDSS/SEGUE efforts are color-selected $in-situ$ surveys, but many of
the VMP stars identified to date are still relatively local.  

The interpretations of (in particular) the kinematics of stars selected by these
two techniques has been the subject of much debate in the literature over the
years. Now we understand why. Stars selected on the basis of high proper motions
favor the identification of members of the outer-halo population, while the
apparent-magnitude-limited prism (and color-selected) techniques favor stars
from the inner-halo population. We presently estimate that no more than 20\% of
halo stars within 4 kpc from the sun are members of the outer-halo population,
but this number has to be confirmed from more detailed modeling. The dominance
of the inner halo continues to at least 10-15 kpc from the Sun; beyond that the
outer-halo population begins to take over. Although there certainly exist
EMP stars that are members of this inner-halo population, they are
completely swamped by the huge numbers of stars near the peak of the
inner-halo MDF around [Fe/H] $= -1.6$. It is surely no accident that the three
UMP (which includes two HMP) stars known at present all are either located at
distances greater than 10 kpc from the Sun, or, in the case of HE~1327-2326 (see
Frebel et al. [8], and also Frebel et al., this volume), exhibit kinematics that
place them firmly in the outer halo.

The clear recognition that outer-halo stars are drawn from a population with 
a lower peak in its MDF ([Fe/H] $= -2.2$) than that of the inner-halo stars
([Fe/H]$ = -1.6$), and the smaller number of stars with [Fe/H] $> -2.0$ that
appear among the outer-halo object,s suggests that one might seek to exploit
these facts in future searches for the most metal-deficient stars. One could, for
example, look to survey fainter stars, in particular distant giants, in order to
reach stars well outside the region where the inner-halo dominates.
Alternatively, one could make use of proper-motion surveys (now and in the
future) to identify stars in the local volume with large components of their
motions in the retrograde direction (or very high $U$ or $W$ motions), since
this appears to strongly favor outer-halo membership. In this manner one might
hope to efficiently identify sufficiently bright UMP and HMP stars that can be
studied at high resolution with current generation 8m-10m class telescopes, and
perhaps one day identify the long sought Mega Metal-Poor (MMP, with [Fe/H] $<
-6.0$, according to [3]) stars.

\begin{theacknowledgments}

Funding for the SDSS and SDSS-II has been provided by the Alfred P. Sloan
Foundation, the Participating Institutions, the National Science Foundation, the
U.S. Department of Energy, the National Aeronautics and Space Administration,
the Japanese Monbukagakusho, the Max Planck Society, and the Higher Education
Funding Council for England. The SDSS Web Site is http://www.sdss.org/.

The SDSS is managed by the Astrophysical Research Consortium for the
Participating Institutions. The Participating Institutions are the American
Museum of Natural History, Astrophysical Institute Potsdam, University of Basel,
University of Cambridge, Case Western Reserve University, University of Chicago,
Drexel University, Fermilab, the Institute for Advanced Study, the Japan
Participation Group, Johns Hopkins University, the Joint Institute for Nuclear
Astrophysics, the Kavli Institute for Particle Astrophysics and Cosmology, the
Korean Scientist Group, the Chinese Academy of Sciences (LAMOST), Los Alamos
National Laboratory, the Max-Planck-Institute for Astronomy (MPIA), the
Max-Planck-Institute for Astrophysics (MPA), New Mexico State University, Ohio
State University, University of Pittsburgh, University of Portsmouth, Princeton
University, the United States Naval Observatory, and the University of
Washington.

T.~C.~B acknowledges support from grants AST 04-06784,  AST 07-07776, and PHY
02-16783, Physics Frontier Centers/JINA: Joint Institute for Nuclear
Astrophysics, awarded by the U.S. National Science Foundation.  D.~C. is
grateful to JINA for support of her long-term visitor status at Michigan State
University, where this analysis took place.

\end{theacknowledgments}

\bibliographystyle{aipprocl} 

\end{document}